\documentclass[sigconf,screen,nonacm]{acmart}

\usepackage{booktabs}
\usepackage{adjustbox}

\AtBeginDocument{%
  }

\begin{document}

\title{On-Premise AIOps Infrastructure for a Software Editor SME: An Experience Report}

\author{Anes Bendimerad}
\affiliation{%
  \institution{Infologic R\&D}
  \streetaddress{99 avenue de Lyon}
  \city{26500 Bourg-Lès-Valence}
  \country{France}
}
\email{abe@infologic.fr}

\author{Youcef Remil}
\affiliation{%
  \institution{Infologic R\&D}
  \streetaddress{99 avenue de Lyon}
  \city{26500 Bourg-Lès-Valence}
  \country{France}
}
\email{yre@infologic.fr}

\author{Romain Mathonat}
\affiliation{%
  \institution{Infologic R\&D}
  \streetaddress{99 avenue de Lyon}
  \city{26500 Bourg-Lès-Valence}
  \country{France}
}
\email{rma@infologic.fr}

\author{Mehdi Kaytoue}
\affiliation{%
  \institution{Infologic R\&D}
  \streetaddress{99 avenue de Lyon}
  \city{26500 Bourg-Lès-Valence}
  \country{France}
}
\email{mka@infologic.fr}

\begin{abstract}
Information Technology has become a critical component in various industries, leading to an increased focus on software maintenance and monitoring. With the complexities of modern software systems, traditional maintenance approaches have become insufficient. The concept of AIOps has emerged to enhance predictive maintenance using Big Data and Machine Learning capabilities.  However, exploiting AIOps requires addressing several challenges related to the complexity of data and incident management. Commercial solutions exist, but they may not be suitable for certain companies due to high costs, data governance issues, and limitations in covering private software. This paper investigates the feasibility of implementing on-premise AIOps solutions by leveraging open-source tools. We introduce a comprehensive AIOps infrastructure that we have successfully deployed in our company, and we provide the rationale behind different choices that we made to build its various components. Particularly, we provide insights into our approach and criteria for selecting a data management system and we explain its integration. Our experience can be beneficial for companies seeking to internally manage their software maintenance processes with a modern AIOps approach.
\end{abstract}



\maketitle

\begin{figure*}[t]
	\centering
	\includegraphics[width=0.9\linewidth]{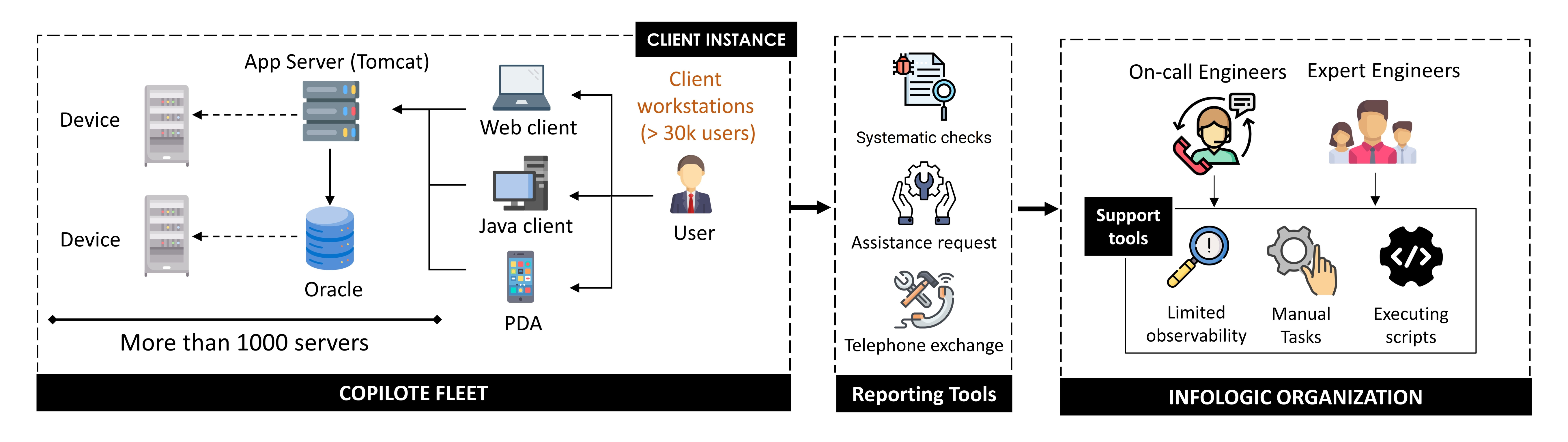}
    \captionsetup{font=footnotesize}
	\caption{A simplified overview of former maintenance {\sc Infologic} system before introducing AIOps and Predictive Maintenance.}%
	\label{fig:old_maintenance}
\end{figure*}

\section{Introduction} \label{sec:introduction}

In today's world, Information Technology (IT) has become a critical element for automation of business process across many industries. Specialized Information Systems and Enterprise Resource Planning (ERP) solutions are widely used by organizations to manage their internal operations, customer and supplier relationships, and Industry 4.0 factories that rely heavily on monitoring and observability. {\sc Infologic}~\cite{infologiccopilote} is one of France's leading providers of ERP solutions for the agri-food, health nutrition, and cosmetic sectors. It has established a significant reputation over its 40 years in the market with a plethora of provided services and modules. {\sc Infologic} also offers its customers  continuous consultation, and proactive maintenance support. Its ERP system is currently deployed by hundreds of food industries in France, each of which has at least one server that is integral to their business and any disruption can result in critical losses. Therefore, it is crucial to ensure high availability and excellent maintenance for this ERP and its infrastructure. 

Software maintenance and monitoring have garnered a notable interest among companies, owing to their increased dependence on software. Meanwhile, the maintenance of such systems is becoming increasingly complex as they always involve more interrelated software and business components, especially with the rise of microservices and virtualization. It follows that traditional maintenance approaches became obsolete and do not scale to today's large systems. These approaches focus on manually performing arduous tasks such as manually logging into devices, executing scripts for checking health status, resolving anomalies on a repetitive basis. Such setting triggers the need for autonomic and self-managing computing systems to address the key reasons for failures and to improve the efficiency and quality of IT services \cite{bogatinovski2021artificial,farshchi2018metric,ren2019time}.

In this context, the term AIOps has been introduced by Gartner~\cite{prasad2018market} in 2017 to address the ITOps (IT Operations) challenges with a data-driven approach that exploits AI. AIOps platforms combine big data and machine learning functionalities to intelligently improve, strengthen and automate a wide range of IT operations, from monitoring to incident management. However, building an effective AIOps solution requires addressing several challenges. Firstly, such solution must be able to handle large and diverse amount of data generated by IT, including metrics, logs, and network traffic. This involves continuously ingesting vast volumes of data from different sources while efficiently storing them to limit the cost of the infrastructure. The AIOps solution should also facilitate data access and efficient querying. Secondly, such solution must enable the monitoring of numerous applications and components while also being easily expandable to include new ones. Thirdly, it needs to simplify the integration of ML techniques into the incident management system, such as applying an anomaly detection technique~\cite{nedelkoski2019anomaly}, or an ML model for incident triage~\cite{DBLP:conf/icse/0003HLXZHGXDZ19,DBLP:conf/kbse/0003HL0HGXDZ19,DBLP:conf/dsaa/RemilBPRK21}. This necessitates being equipped with comprehensive MLOps tools to streamline data preparation, model training and testing, as well as model deployment and monitoring. Lastly, it is important to ensure that the solution is highly accessible and intuitive to facilitate the transition of engineers from traditional approaches to AIOps. In fact, experience confirms that shifting the mindset and work habits within ITOps can pose a significant challenge~\cite{dang2019aiops}.

Commercial solutions have been proposed to address these challenges by providing AIOps features such as automatic discovery, flexible data visualization, intelligent alert generation and incident diagnosis, automatic incident triage and resolution. Few examples of popular platforms are Dynatrace~\cite{dynatrace}, Splunk~\cite{splunk}, Datadog~\cite{datadog}, New Relic~\cite{newrelic}, IBM Instana~\cite{instana}, and PagerDuty~\cite{pagerduty}. These platforms have witnessed an impressive success, with many of them already exceeding a billion dollars in annual revenue. While their effectiveness is undeniable, their cost can be prohibitive depending on the context, and their pricing structures often challenging to manage. Particularly at {\sc Infologic}, with over a thousand servers hosting various applications, integrating a commercial solution can prove to be very expensive. Their cost is determined by multiple parameters, such as the number of monitored servers, the required storage size, the number of stored event logs. For example, our estimation of the cost for one of the aforementioned solutions ranges between \$500k and \$1 million per year to fully integrate it as a monitoring tool for {\sc Infologic}. Additionally, while commercial AIOps solutions can monitor a broad range of popular applications, they do not cover private software such as {\sc Infologic}'s ERP, particularly its functional and business layer. Thus, integrating a commercial AIOps solution in our context would still require significant development efforts. Moreover, the use of such solutions may pose data governance issues as it often involves delegating data control to an external entity, which some of {\sc Infologic}'s clients may not be comfortable with, thus limiting our adoption of such tools.

Meanwhile, there has been a surge of open-source software solutions. Among them, numerous projects help to deal with some challenges related to AIOps, while not directly providing a complete AIOps platform. For instance, several open-source agents, such as Telegraf from InfluxData~\cite{telegraf} and Beats from the ELK stack~\cite{beats}, enable seamless collection of data from applications. Also, open-source engines have been created to efficiently manage large volumes of data. Some of these solutions focus on specific types, such as metrics in the case of InfluxDB and TimescaleDB, while others are more generic, like ClickHouse and Elasticsearch. Therefore, it is becoming increasingly feasible for medium-size software companies, like ours, to develop in-house and modern AIOps platforms by combining and building on top of open-source components.

\noindent \textbf{Contributions.} We have thoroughly explored this possibility and investigated the different open source solutions and architectures that we can exploit to deploy an On-Premise AIOps system. In this paper, we share our experience by presenting our end-to-end on-premise infrastructure, while explaining the methodologies that we followed to make various decisions. Particularly, we provide insights into our approach and criteria for selecting a data management system and we explain how we integrated it. Although we have deployed this platform on our own servers, it can also be implemented on an IaaS Cloud solution which provides physical servers and manages them for its clients.

\noindent \textbf{Outline.} Sec~\ref{sec:background} provides an overview of {\sc Infologic} and its ERP software called Copilote, along with an introduction to AIOps capabilities. In Section~\ref{sec:challenges}, we delve into the maintenance and monitoring challenges faced in our software, as well as the obstacles encountered when deploying an on-premise AIOps solution. To address effectively these challenges, we establish a list of considered criteria that are provided in Sec~\ref{sec:criteria}.  Sec~\ref{sec:infrastructure} introduces our proposed AIOps infrastructure and elucidates the various choices we made. Finally,
we present discussions and conclusions in Sec~\ref{sec:discussion}.

\section{Background} \label{sec:background}
\subsection{Infologic and Copilote}\label{subsec:infologic}
{\sc Infologic}  is a leading provider of ERP solutions for the agri-food, health nutrition, and cosmetic sectors in France. Its flagship product, Copilote, is an enterprise resource planning software designed to optimize and automate a large panel of business processes, including sales tracking, supply chain management, and customer relations. {\sc Infologic} manages the complete Copilote implementation process, which includes integrating the software for the client, providing training, and maintaining the ERP instances and infrastructure in operation.  This infrastructure comprises many dependent components and applications such as a Tomcat server, Oracle or Postgres databases, virtual machines, backup systems, and more. Currently, this ERP system is deployed by hundreds of food industries, ranging from small companies to large corporations.
As the proper functioning of these businesses depends heavily on the reliable performance and accessibility of the ERP, it is crucial to ensure high availability and excellent maintenance for Copilote.

Before 2019, {\sc Infologic}'s service was mainly focused on systematic and corrective maintenance, that is, apply time-based checks and solving incident tickets that are created by clients. Fig~\ref{fig:old_maintenance} depicts the former maintenance system before introducing AIOps. The user can interact with the ERP through a web client, a Java client, or a PDA (Personal Digital Assistant). During these interactions, she may encounter some errors, outages, or difficulties of usage, for which she would submit an incident or assistance ticket for {\sc Infologic}. OCEs (On-Call engineers) are the first to receive these tickets. If feasible, they can address and resolve them directly, otherwise, they route them manually to expert engineers. The scope of predictive maintenance was constrained due to a notable absence of tools and limited system observability. For instance, the lack of centralized monitoring data necessitated connecting to servers to access component-specific observations. Additionally, data retention was typically limited to one week in most cases, as storing more data using traditional database systems could incur high costs due to the high volume and velocity of the data.

Although this former approach may have been sufficient a decade ago, it no longer scales as the number of clients grows. In fact, the size of the environment production fleet has exceeded a thousand servers, each of them handling various applications. {\sc Infologic} is confronted with an important challenge of enhancing operational efficiency, all while ensuring optimal customer satisfaction levels are upheld. Since 2019, {\sc Infologic} has invested a lot on shifting the approach and focusing more on Predictive Maintenance rather than Corrective Maintenance. This can be achieved by following a data-driven philosophy where data is at the heart of any operation, starting from observability to automated recovery. 

\subsection{AIOps}\label{subsec:aiops}

\begin{figure}[t]
	\centering
	\includegraphics[width=0.93\linewidth]{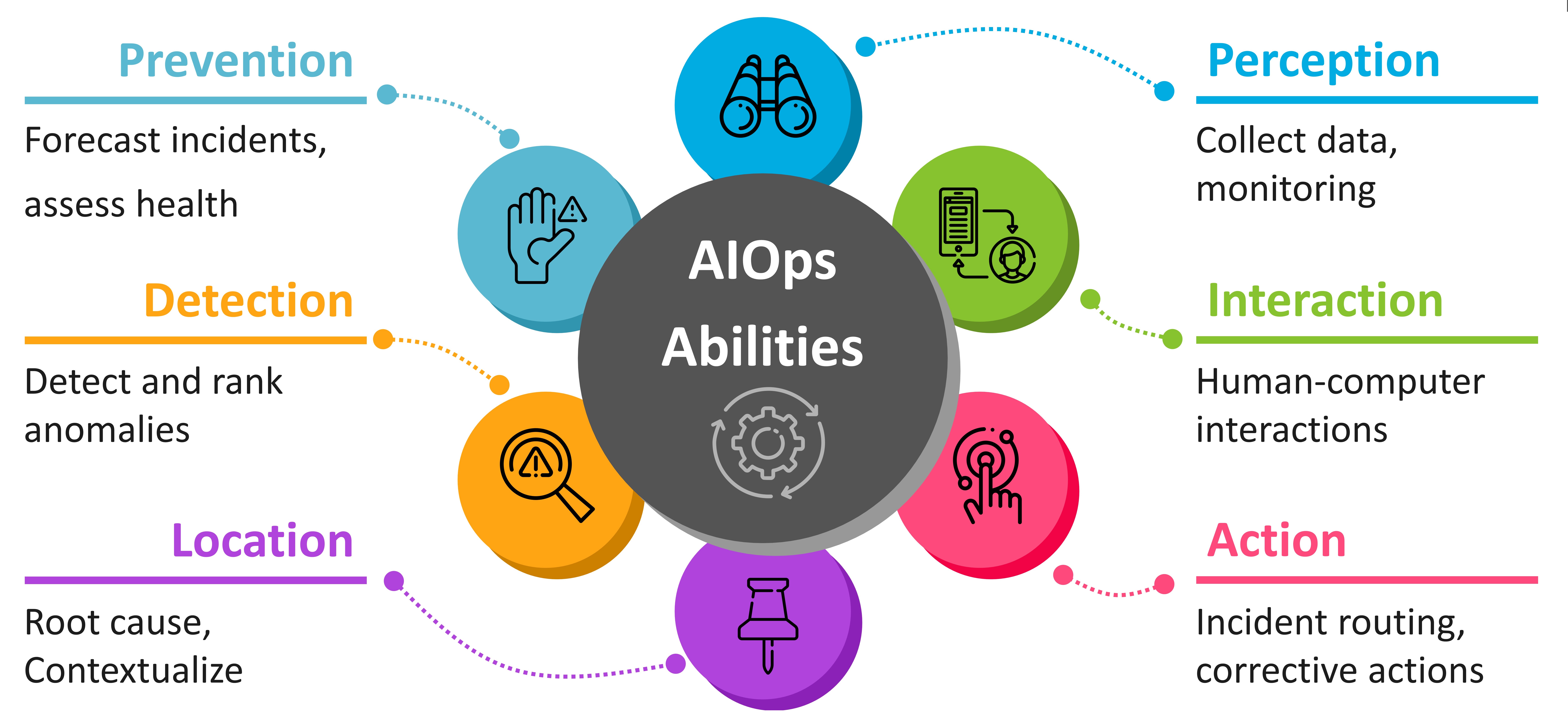}
    \captionsetup{font=footnotesize}
	\caption[Main AIOps Features.]{Main AIOps Features}%
	\label{fig:aiops_abilities}
\end{figure}

The challenges encountered by {\sc Infologic} have sparked our interest toward replacing  manual maintenance routines with a unified and intelligent approach to conduct maintenance incidents from creation through diagnosis to resolution. The ultimate objective is to automate as many of the associated tasks as feasible, with a view to optimize the so called MTTD (Mean Time To Detect) and MTTR (Mean Time To Resolve). This approach is referred to as AIOps, which stands for AI for Operating Systems. AIOps leverages big data and machine learning to intelligently automate a wide range of IT and maintenance operations, and to accelerate the identification and resolution of IT issues and outages~\cite{prasad2018market, dang2019aiops}. According to research~\cite{shen2020evolving, prasad2018market, rijal2022aiops}, a prototypical AIOps system ought to encompass six fundamental abilities that cover diverse tasks, as shown in Figure~\ref{fig:aiops_abilities} and detailed below.

\noindent \textbf{Perception.} It corresponds to the capacity of gathering heterogeneous data types, including logs, metrics, and incident tickets, from multiple sources. The ingestion process needs to accommodate both real-time streaming and historical data analysis. Additionally, powerful data visualization and querying are necessary to ensure optimal response times when accessing the data.

\noindent \textbf{Prevention.} This process involves anticipating potential failure scenarios and forecasting high-severity outages before they occur in the system, utilizing simple alert rules, statistical models and predictive machine learning algorithms. 

\noindent \textbf{Detection.} If errors or performance issues occur, the system must detect related anomalies. This involves analyzing extensive historical data to spot abnormal occurrences, while reducing false alarms and redundant events.

\noindent \textbf{Location.} The objective of this process is to identify and analyze potential root causes responsible for errors. This involves conducting a causality and correlation study, as well as providing tools that well characterize the context of those errors. 

\noindent \textbf{Action.} Once the root cause of a problem is identified, the next step is to act to solve it. The AIOps platform must provide simplified actions based on the current context and past solutions outlined in the prescriptive maintenance protocol. Some actions can even be triggered automatically once a corresponding incident is detected.

\noindent \textbf{Interaction.} AIOps solutions should enable interactive analysis between the intelligence provided by integrated models and the expertise of users. This includes facilitating communication between different maintenance teams or customers, promoting efficient information sharing and problem resolution. This is achieved through various means, such as automatic incident documentation and summarization, or chatbot agents utilizing NLP techniques.

\section{Challenges} \label{sec:challenges}
While adopting an AIOps solution can deliver significant benefits to our organization, it comes with numerous challenges that need to be carefully considered during the implementation of such solution. 

\noindent \textbf{Human challenge.} One of the main challenges is the difficulty of shifting the mindset of IT professionals to adopt new ways of working~\cite{dang2019aiops}. A notable effort needs to be invested to simplify this transition. Thus, it is important to ensure interpretability and explainability of AIOps solutions to build trust in these approaches~\cite{lyu2021towards}. 

\noindent \textbf{Novelty of AIOps.} AIOps is still a recent and unstructured field~\cite{notaro2021systematic}. Given its novelty and cross-disciplinary nature, AIOps contributions and methods are widely dispersed, lacking standardized taxonomic conventions for data management, as well as clear capabilities and implementation details. As such, discovering and comparing these methods can be a challenging endeavor. Therefore, one should determine the class of methods and tools that are needed given the defined goals, and establish a clear and objective procedure for comparing and selecting the right ones.

\noindent \textbf{Handling large volumes of data.} Data plays a crucial role in different aspects of AIOps. Ensuring high observability and real-time analysis requires efficient and continuous integration of large volumes of data from multiple sources. However, this integration poses several challenges. Foremost, data collection routines must not negatively impact the performance of the monitored applications, such as overwhelming their memory or network resources. Additionally, the data management system used must exhibit high performance in terms of data ingestion, compression, and querying capabilities across various data types to support this process. 

\noindent \textbf{Data quality.} Without effective data quality control, AIOps data may exhibit numerous defects. For example, network failures can lead to missing data, which undermines both data observability and representativeness. Message queues can help to address this issue~\cite{DBLP:journals/access/FuZY21}. Moreover, data related to incident management is largely sourced from human actions, such as incident tickets, corrective actions, and labels, which may be prone to human errors. This can lead to inaccuracies when using this data to train AIOps models. 

\noindent \textbf{Integrating ML models.} AIOps presents challenges in integrating ML solutions. For instance, anomaly detection is a key task in AIOps, but data in this field is extremely imbalanced as anomalies are rare. Moreover, lack of labeled data is a common issue, limiting the scope of applications to unsupervised or semi-supervised methods. Another important fact is that AIOps data is prone to concept drift~\cite{lu2018learning}. Software systems evolve for many reasons, such as the introduction of new components or an increase in the number of users. Any of these changes may impact significantly the characteristics of data. Therefore, it is essential to leverage MLOps tools~\cite{kreuzberger2023machine} to  monitor model quality in production, and update models accordingly.

\noindent \textbf{Integrating expert knowledge.} Although modern ML techniques can derive valuable insights from data, this data may not contain all the requisite knowledge to accurately perform the given task, particularly in AIOps where there is a clear lack of ground truth. Moreover, due to the evolving nature of a software system, certain knowledge and inferences may become obsolete, while new knowledge and ways of dealing with problems emerge. Hence, an AIOps platform should support an interactive process that enables the integration of knowledge from expert engineers and help to continuously improve the quality of algorithms' results. Active Learning~\cite{settles2009active,DBLP:journals/csur/RenXCHLGCW22} can be especially beneficial in this context.

\noindent \textbf{Security and Privacy.} AIOps heavily relies on data, including sensitive information that could pose potential security threats or privacy concerns if not handled properly. An AIOps platform may need to track user interactions with the monitored software, or collect incident tickets containing confidential information about clients. Thus, we must ensure a high standard of security and comply with regulations such as the General Data Protection Regulation (GDPR)~\cite{gdpr}. Privacy is one of the concerns that prompt us to implement an on-premise platform, to have full control over our data.

\begin{figure*}[t]
	\centering
	\includegraphics[width=0.95\linewidth]{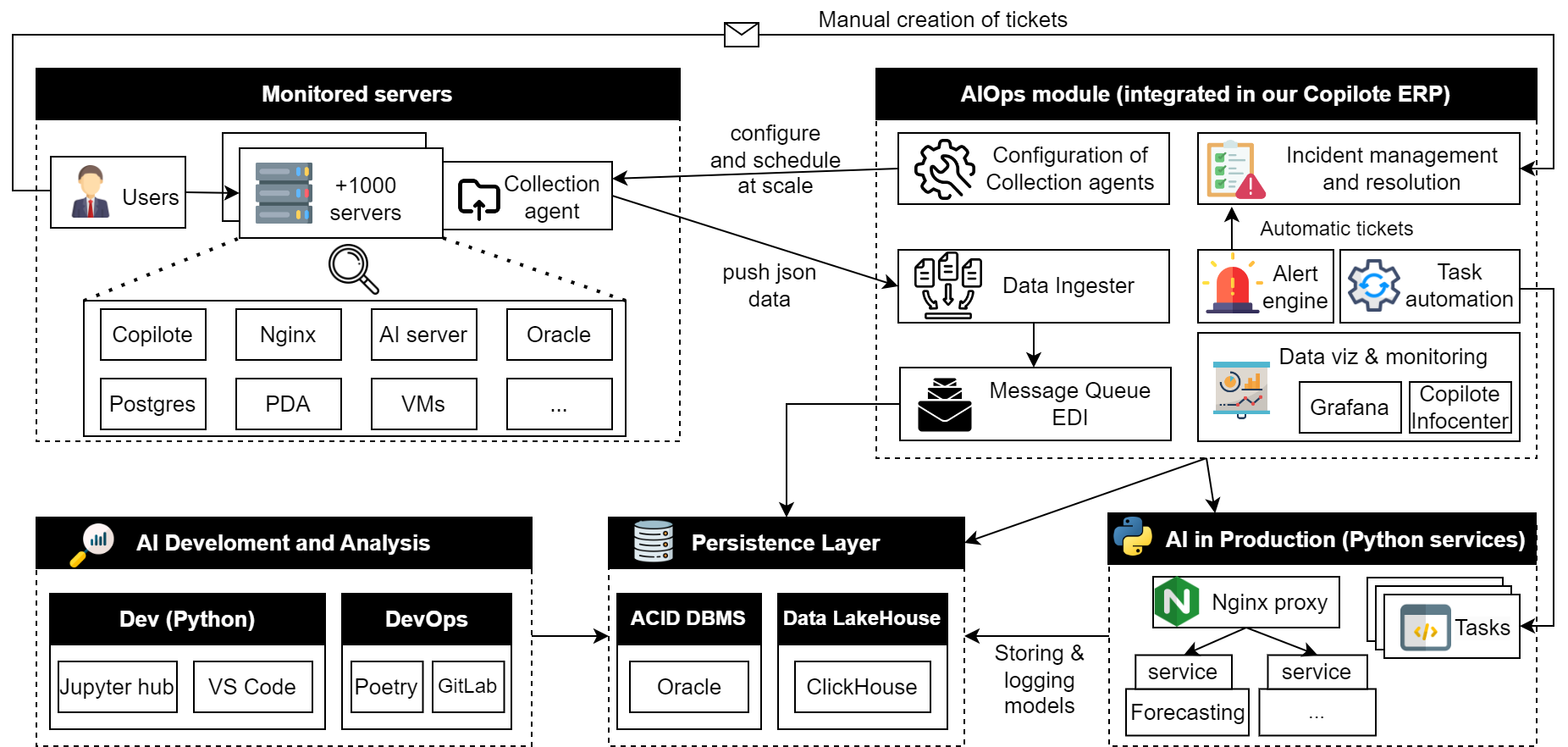}
    \captionsetup{font=footnotesize}
	\caption{On-premise AIOps Infrastructure: Overview and main used tools.}%
	\label{fig:infraAIOps}
\end{figure*}

\section{Criteria of choice} \label{sec:criteria}

During the construction of our AIOps infrastructure, we had to make multiple decisions regarding the selection and configuration of architectural components, as well as choosing between using existing external tools or developing custom solutions. These decisions aimed at providing an AIOps solution that effectively address the challenges we previously identified while ensuring optimal efficiency. To make well-informed choices, we have established a list of objective criteria that guided the implementation of this project. These criteria were tailored to suit the varying circumstances and specific requirements of each component within the architecture. 

\noindent \textbf{License.} This aspect concerns external projects that can be integrated into our infrastructure.  A software solution is subject to a license that defines the terms and conditions of its usage. Several licenses  are recognized as open-source, but some of them are more permissive than others. In our context, we are mainly interested in open-source solutions whose license permits both the usage and commercialization of the software. It should be noted that some solutions, such as Elasticsearch ~\cite{elasticsearch} and MongoDB ~\cite{mongodb}, have licenses that are copyleft. Although we prefer a fully permissive license, such as Apache license or MIT license, we also consider solutions whose open-source usage is limited to the default distribution. 

\noindent \textbf{Notoriety and popularity.} This is a relevant indicator which suggests that a project has been widely tested, validated, used, and accepted by the community. Moreover, popular projects tend to benefit from stability, continuous development, updates, and support. There exist metrics that can be used to estimate a project popularity, such as the overall number of downloads and deployment, the number of GitHub stars and forks, the number of active contributors. Some factors that are relatively subjective can also be considered, such as recommendations of experts.

\noindent \textbf{Cost of implementation.} This includes human and material resources that are required to implement a solution. When considering an external project that is already developed, the costs primarily involve deployment and adaptation to our ERP system. On the other hand, if we are developing a custom component from scratch, the main factor influencing this criterion is the cost of development.

\noindent \textbf{Operational cost.} Since we aim for an on-premise solution, this criterion involves the cost of infrastructure needed to operate the software, alongside the cost of its administration and monitoring. In contrast, in the case of a cloud-hosted solution, the primary consideration lies in the pricing associated with utilizing the service.

\noindent \textbf{Flexibility and Adaptability to internal context.} This criterion assesses the ease of integrating and adapting a tool to the existing technology stack and adhering to our company's standards. In our specific context, this factor holds significant importance since the AIOps infrastructure must seamlessly integrate with the existing ERP, which possesses its own characteristics and utilizes internal tools that can potentially be leveraged in the AIOps project. Furthermore, the monitoring of such a proprietary software necessarily implies specific developments to cover all its scope from technical to business layer. 

\noindent \textbf{Ease of usage.}  This is a crucial criterion as it heavily affects the adoption rate of a tool, which in turn, impacts the success probability of the project. For instance, when evaluating database systems,  one key factor that influences this criterion is the query language used by the solution. Systems that use SQL such as ClickHouse and TimescaleDB are generally easier to adopt than those that employ specific query languages such as Elasticsearch and InfluxDB. 

\noindent \textbf{Security and privacy.} Given that AIOps relies on critical data sources and applications, we need to ensure that all the architecture has strong security standards, especially when integrating external projects. Evaluating this aspect when integrating open-source tools into the project can be accomplished through various criteria, including code reviews and audits, community engagement on good practices, security features and their documentations, and security track record.

\noindent \textbf{Limiting the number of tools.} It is important to keep in mind that when the number of distinct tools employed in the infrastructure increases, so does the complexity and cost associated with its maintenance.  In fact, each tool may require additional specific skills and engender more complex dependencies within the architecture.

\noindent \textbf{Performance and scalability.} Performance concerns the efficiency of a tool in executing its assigned task. It can be measured by the resources required to accomplish these tasks, such as time, memory, CPU, network resources. On the other hand, scalability corresponds to the capacity of the solution to scale to larger workloads, and this refers to both vertical and horizontal scalability. 

\begin{figure*}[t]
	\centering
	\includegraphics[width=0.99\linewidth]{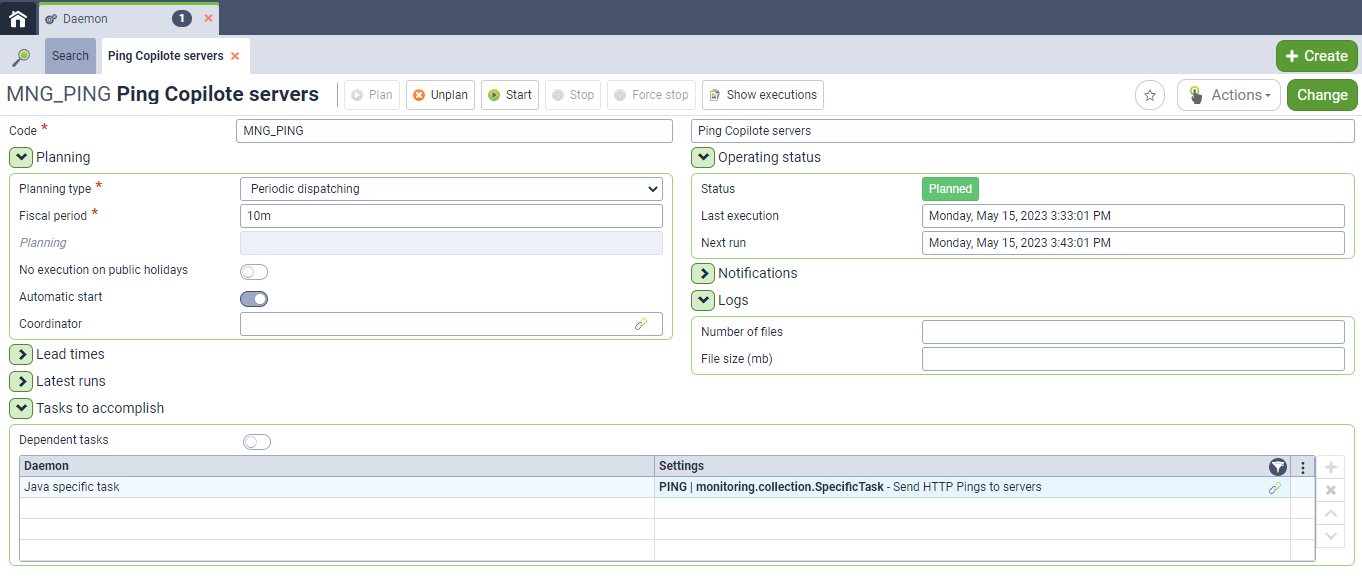}
    \captionsetup{font=footnotesize}
	\caption{Agent configuration tool: Example of configuration of a task that pings Copilote servers every 10 minutes.}%
	\label{fig:config_collect}
\end{figure*}

\section{On premise AIOps Infrastructure} \label{sec:infrastructure}

Fig~\ref{fig:infraAIOps} provides an overview of the infrastructure and highlights the key tools used to implement different aspects of AIOps. Our monitored infrastructure comprises over a thousand servers that host the Copilote ERP and other applications that interact with it. To extract monitoring data, a collection agent is installed on each server and configured at scale using our new AIOps module, which is fully integrated into Copilote. This module ingests data from collection agents and transmit them to a Message Queue. Then, these data are stored in a relational ACID DBMS or a Data LakeHouse~\cite{DBLP:conf/cidr/Zaharia0XA21,harby2022data}, depending on the use case. Our AIOps module includes visualization tools to explore these data and create dashboards that facilitate monitoring. Moreover, a collaborative platform has been implemented to enable data scientists to analyze and experiment with collected data and develop AI algorithms. Effective algorithms can be then deployed in the AI in Production infrastructure as a web-service that serves REST queries on demand or a Python task that is orchestrated by the Copilote AIOps module. We have also introduced an internal alerting engine that can generate alerts using simple rules or advanced Machine Learning techniques, by interacting with the AI services and tasks. This engine can be configured to automatically create incident tickets that will be then handled by the incident management and resolution node. It is also possible to manually create an incident ticket. In what follows, we describe in detail these main infrastructure components.

\subsection{Collection agent} 
This is a lightweight application that is installed on monitored servers, and that can interact with various server components to gather their health status, performance metrics, logs, and more. Notable components that we need to monitor include our ERP software, related databases such as Oracle or PostgreSQL, operating systems, virtual machines, and more. The most critical component to monitor is the ERP software, including its technical, applicative, functional, and business layers. There exist popular open-source collection agents that can be considered in this study. For instance, InfluxData offers Telegraf~\cite{telegraf} which is able to connect with more than 300 popular applications. The ELK stack provides the Beats suit~\cite{beats} to collect and send metrics to Elasticsearch. Fluentd~\cite{fluentd} is another agent that can collect events and send them to various types of destinations such as files or DBMS. 

These tools offer numerous advantages, being open-source with substantial popularity indicated by at least 10k GitHub stars and thousands of forks. Moreover, their deployment and configuration is simple. However, they do not provide tools to gather specific data related to our ERP since it is a proprietary software, which represents a significant portion of our AIOps monitoring scope. Additionally, leveraging these tools does not allow us to directly utilize our internal data manipulation libraries, a factor that could simplify our work and enhance maintainability. Therefore, we built upon existing tools and libraries to implement our own collection agent. This choice also enables us to tailor the agent to specific network constraints that we have at Infologic. 

For each installed agent, we can schedule a list of collection tasks that are continuously executed. A task can be a shell script, a Java or Python code snippets, an SQL query, or an HTTP Rest query. An example of a simple shell task that is executed every 10 minutes with the agent to get the total number of processes in the Operating System is ``\texttt{ps -ef --no-headers | wc -l}''. The result of a task is then sent to our AIOps module that consolidates and stores the data. To address potential network failures, our collection agent is able to buffer the data collected for a certain duration and retry sending later. This ensures that our AIOps monitoring remains reliable and robust even in the face of potential network disruptions.

\subsection{Agent configuration tool}
We have introduced an agent configuration node within the AIOps module of our ERP. This node is used to perform the installation and configuration at scale of data collection agents on the monitored servers. Such configuration tool is crucial since we are applying an agent-based monitoring whereas collection agents need to be installed, configured, and continuously updated on a thousand servers. This tool makes it possible to flexibly configure a data collection task on a specific subset of servers. For example, to monitor database tables, we need to configure the collection of their sizes and number of rows on only DBMS servers. This is made possible thanks to the agent configuration tool. 
In Fig~\ref{fig:config_collect}, we show an example of a task that is scheduled using our agent configuration tool. This task performs a basic ping on all our servers to check whether they are alive. The configuration tool offers buttons for task planning, unplanning, and other related actions.  In the "Planning" section, we can observe that the task is scheduled to run every 10 minutes. Additionally, this tool provides information and logs about the previous task executions, facilitating monitoring and troubleshooting.  

We developed our own configuration tool for the same reasons that motivated us to create our own agent. Moreover, this configuration tool was built by extending our ERP task management module, which is widely used in Copilote to schedule tasks in various business modules such as sales and supply chains. Utilizing this module allowed us to leverage our strong internal expertise for its development and ensured direct compatibility with our monitoring and alerting system, which simplified its adoption. Here again, there are open-source solutions that can be used in this context, such as Ansible~\cite{ansible} from Red~Hat. Some alternative tools provide an agentless approach to collect data, which eliminates the need of installing agents on monitored servers. Such approach generally requires less effort to set up a generic monitoring, it is known to be limited as it lacks the same level of interaction with applications like agent-based monitoring. In~\cite{pandey2020analogy}, an in-depth comparison of pros and cons of agent-based and agentless monitoring is detailed.

\begin{figure}[t]
	\centering
	\includegraphics[width=0.99\linewidth]{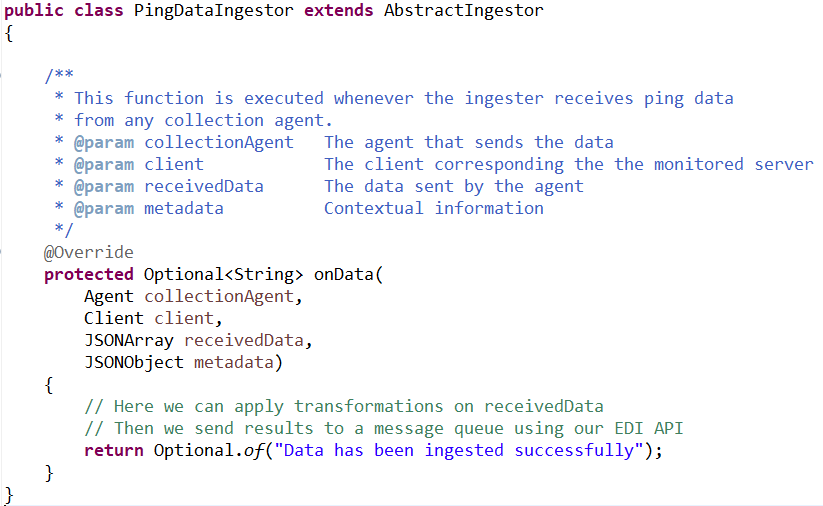}
    \captionsetup{font=footnotesize}
	\caption{An example of class that can be used to perform ingestion of ping data using Java internal APIs.}%
	\label{fig:ingester}
\end{figure}

\subsection{Data ingester} 
Our AIOps module provides a data ingester that receives monitoring data from collection agents.
The JSON format has been mainly used to transmit data, as it is one of the simplest and standard data format. Transmitted data are automatically compressed using the popular GZIP algorithm. Thanks to this ingester, data transformation and augmentation can be performed before sending it to the message queue. In Fig~\ref{fig:ingester}, we show the Java class and the method \texttt{onData(...)} which can be used to ingest data that we receive from collection agents. This method is executed when an agent sends new data provided by the variable \texttt{receivedData}. At this level, we can flexibly perform data transformation and then send the result to a message queue using our internal EDI (Electronic Data Interchange) API. Our framework also provides a generic data ingester that simply receives the data and sends it to a specified message queue without modifying it. 

A notable open-source alternative to ingest data is Logstash~\cite{logstach}. While it has been mainly developed to ingest data coming from beats and going to Elasticsearch, it is now compliant with other databases and agents like Telegraf. Since the implementation of our data ingester had a limited and controlled cost, we opted for this solution to achieve full integration with our software to avoid adding an unnecessary dependency on an additional tool like Logstash.

\subsection{Message queue system}

\begin{figure*}[t]
	\centering
	\includegraphics[width=0.99\linewidth]{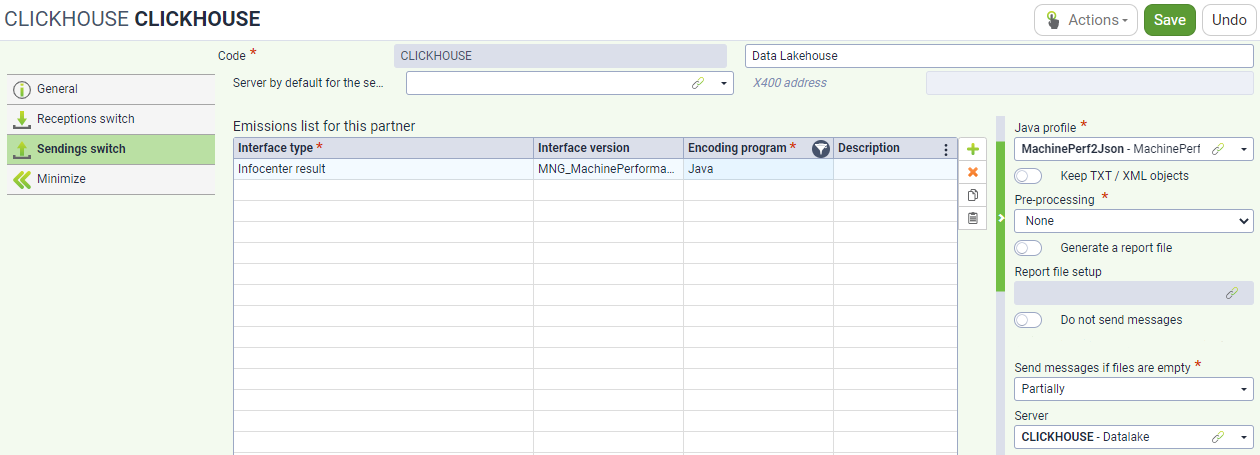}
    \captionsetup{font=footnotesize}
	\caption{An EDI interface used to persist data on ClickHouse. In this simplified example, it consumes only one topic related to machine performance data.}%
	\label{fig:clickhouse_edi}
\end{figure*}

A message queue system serves as an interface between the data ingester and data consumers, offering several benefits. Firstly, it ensures data durability, meaning data remains in the message queue until consumed by all the consumers that require it. Additionally, it simplifies the consumption of data by multiple destinations at once, as certain data needs to be stored in several database systems. Furthermore, asynchronous communication via the message queue notably improves ingestion performance by allowing the ingester to send data without waiting for it to be processed. 

Our ERP already includes a built-in message queue module founded on the EDI standard~\cite{DBLP:journals/jmis/PremkumarRN94}, which is extensively used across various business modules of Copilote. Hence, we have leveraged this fully integrated feature for our AIOps module. An example of the EDI interface used for sending data to ClickHouse is illustrated in Fig~\ref{fig:clickhouse_edi}. In  ``Emissions list for this partner'', we define a list of ``topics'' that are consumed by this interface and stored on ClickHouse. These same topics can be consumed by other EDI interfaces as well, allowing them to be sent to additional destinations when needed. In this example, we have also attached a Java profile (displayed on the right-hand side of the screen), which is a code snippet that enables final data transformation specifically related to ClickHouse.

Had we not had the ready-to-use EDI module, we could have utilized one of the open-source message queue platforms such as Apache Kafka~\cite{kafka}, RabbitMQ~\cite{rabbitmq}, or RocketMQ~\cite{rocketmq}. These systems vary in design and characteristics, suiting different project requirements. A thorough comparison of popular message queuing systems is provided in~\cite{DBLP:journals/access/FuZY21}. This paper highlights the distinct features of each system and summarizes their best-suited cases.

\subsection{Persistence layer}
Consumers from the message queue send data to the persistence layer for storage. Due to significant variations in data characteristics and significance, we could not rely on a ``one size fits all'' database in this context, as also supported by similar studies~\cite{shen2020evolving,yeruva2023monitoring}. For instance, some data related to the general characteristics of servers and their clients require strong consistency but are of limited size. In this case, an ACID-compliant relational DBMS like Oracle is suitable. However, such a database system is not well-suited for storing thousands of performance metrics collected from each server at high frequency and retained for extended periods. For example, to monitor CPU usage alone, we collect dozens of metrics from a thousand servers at a frequency of 0.1 hertz. This data, stored for two years, results in a table with six billion rows and multiple columns. Log data presents an even greater challenge, with every web service call generating multiple lines of logs, considering over 100 million web service calls collected daily. We have devoted significant effort to identifying the appropriate data management tools for these wide variety and volume data sets. In the first version of our AIOps module, we used InfluxDB for time series data and Elasticsearch for textual data. However, we recently shifted to ClickHouse, which is highly efficient in managing various types of data, including metrics and text, and is easier to use than InfluxDB and Elasticsearch, thanks to its SQL syntax. Recent studies have shown the superiority of ClickHouse on online analytics~\cite{mostafa2022scits,clickbench} compared to competitors. Therefore, 
Several companies, such as Contentsquare~\cite{contentsquareClickHouse}, eBay~\cite{ebayClickHouse}, Cloudflare~\cite{cloudfareClickHouse}, and Disney+~\cite{disneyClickHouse}, have migrated to ClickHouse from other solutions. 

To verify that ClickHouse is suitable for our context, we conducted a comparative study whose results led us to migrate our Data LakeHouse to this solution. This choice also simplifies administration and reduces costs by consolidating to one tool instead of two (InfluxDB and Elasticsearch). We have compared it with three other popular open source solutions which are InfluxDB, Elasticsearch, and TimescaleDB. In what follows, we detail further the position of each solution given specific criteria.

\noindent \textbf{License.} The four solutions we have considered all offer licenses that at least allow unrestricted usage of their default distributions. However, the extent of features included in these distributions varies between the solutions. For instance, the open-source version of InfluxDB lacks horizontal scalability, limiting its usage to a single node, while the three other alternatives can be freely used in distributed settings. Among them, ClickHouse and TimescaleDB have the most permissive licenses, being covered by Apache License 2.0.

\begin{figure}
	\centering
	\includegraphics[width=0.95\linewidth]{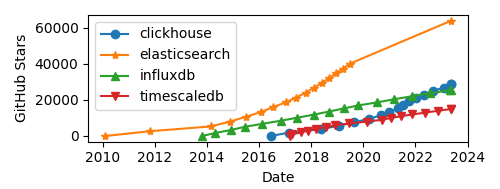}
    \captionsetup{font=footnotesize}
	\caption{Evolution of number of GitHub stars for Elasticsearch, ClickHouse, InfluxDB, and TimescaleDB.}%
	\label{fig:stars_history}
\end{figure}

\noindent \textbf{Notoriety and popularity.} All the four solutions have established a strong popularity over the open-source community. To compare their GitHub repositories' activity, we provide interesting metrics in Tab~\ref{tab:popularityDatabase}. The repository with the highest number of stars is Elasticsearch. Fig \ref{fig:stars_history} illustrates the evolution of stars over time, showing a notable increase for ClickHouse, which has recently surpassed InfluxDB. Although all four repositories demonstrate considerable activity across the other metrics listed in Tab~\ref{tab:popularityDatabase}, TimescaleDB has relatively smaller values. In addition, ClickHouse stands out with a significantly higher number of commits in the past year, indicating a faster project evolution rate. These metrics provide insights into the popularity of these projects, but they do not necessarily imply that one project is universally more interesting than another. 

\begin{table}[]
\captionsetup{font=footnotesize}
\caption{Metrics comparing GitHub repositories of the four DBMS considered in our study.}%
	\label{tab:popularityDatabase}
 \begin{adjustbox}{width=0.95\columnwidth,center}
\begin{tabular}{@{}lllll@{}}
\toprule
\textbf{Solution} & \textbf{Stars} & \textbf{Forks} & \textbf{Contributors} & \textbf{Last year commits} \\ \midrule
ClickHouse        & 29k            & 5.7k           & 1.2k                  & 26k                                 \\
Elasticsearch     & 64k            & 23k            & 1.8k                  & 4.5k                                \\
InfluxDB          & 27k            & 3.5k           & 454                   & 4.3k                                \\
TimescaleDB       & 15k            & 793            & 83                    & 700                                 \\ \bottomrule
\end{tabular}
\end{adjustbox}
\end{table}

\noindent \textbf{Ease of usage and adaptability.} This is an important factor that greatly influences a project's success. Our experience has shown that it is extremely helpful when the used query language is SQL, as it is widely popular. InfluxDB and Elasticsearch employ their own specific languages for data querying, which has hindered their adoption at Infologic. In contrast, ClickHouse, with its SQL-based query language, has witnessed a faster adoption within our company. Another aspect for evaluating the ease of usage and adaptability of a solution is the availability of libraries and tools that facilitate its integration with other projects. For instance, integrating a database source on a visualization tool like Grafana. All the four studied solutions benefit from extensive library support that ease their integration on different projects.

\noindent \textbf{Performance.} Numerous empirical studies have been performed to compare DBMS solutions for online analytics on large volumes of data. \citet{mostafa2022scits} provide a thorough study that evaluates ClickHouse, InfluxDB, TimescaleDB, and PostgreSQL in handling time series data. It compares their  ingestion and storage capacity, and their performance on executing different types of queries. This study shows the superiority of ClickHouse on handling metrics compared to other solutions. While it also indicates satisfactory performance of InfluxDB, this study has not evaluated it on datasets with high time series cardinality. In fact, it is known that the Achilles heel of InfluxDB is when the number of time series exceeds a few millions, whereas the performance of the open-source InfluxDB version becomes poor. Furthermore, ClickBench~\cite{clickbench} is a wide benchmark that showcases ClickHouse as one of the top performers in online analytics compared to over thirty other solutions, including Elasticsearch, TimescaleDB, and several paid cloud solutions. These findings are reinforced by the experiences of various companies. For example, Contentsquare~\cite{contentsquareClickHouse} migrated its infrastructure from Elasticsearch to ClickHouse, reducing infrastructure costs by eleven times while storing six times more data and achieving a tenfold improvement in query execution speed for 99\% of queries. Similarly, eBay~\cite{ebayClickHouse} reduced its infrastructure footprint by 90\% after migrating from Druid to ClickHouse. Cloudflare~\cite{cloudfareClickHouse} improved its performance by migrating from CitusDB to ClickHouse, and Disney+~\cite{disneyClickHouse} chose ClickHouse over Elasticsearch, Apache Flink, and Hadoop after thorough evaluation.

\noindent \textbf{Scope of usage.} ClickHouse and Elasticsearch are generic databases capable of handling various types of data, including time series, text, logs, and unstructured data in JSON format.   In contrast, InfluxDB and TimescaleDB are specifically designed for time series data. Despite this specialization, our performance study did not demonstrate superior performance of the specialized solutions. Therefore, it is not advantageous for us to invest in a dedicated time series database, which would increase the complexity of our system.

\subsection{Data visualization and monitoring}
\begin{figure}[t]
	\centering
	\includegraphics[width=0.99\linewidth]{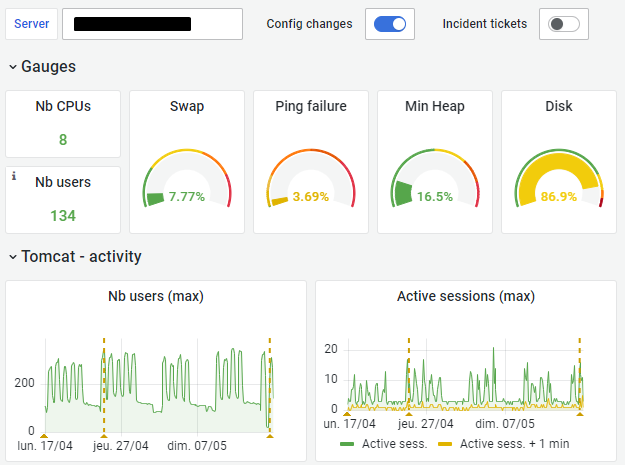}
    \captionsetup{font=footnotesize}
	\caption{Example of a Grafana dashboard that visualizes data loaded from both ClickHouse and Oracle.}%
	\label{fig:grafana_dashboard}
\end{figure}

Efficient data visualization tools are a critical component of a successful AIOps project. These tools must offer a user-friendly interface and provide a variety of visualization panels that are suitable for monitoring, such as time series, gauges, dynamic topologies. These tools help IT professionals to monitor the infrastructure, diagnose problems, identify component relationships, navigate to root causes, and make informed decisions. Dashboards, which are pre-configured sets of visualization panels that succinctly represent data in a single view, are one of the primary uses of these tools. Our AIOps module integrates two visualization tools. The first is Copilote infocenter, our BI tool, which is fully integrated into our software and simplifies manipulation of data stored in the Oracle database of the ERP using a drag-and-drop approach. However, this tool currently cannot interact with external data sources such as ClickHouse. To overcome this limitation, we also use Grafana~\cite{grafana}, an open-source tool that allows us to query data from different sources including Oracle and ClickHouse. Fig~\ref{fig:grafana_dashboard} shows a Grafana dashboard including data loaded from both ClickHouse and Oracle. Another notable advantage of Grafana dashboards and panels is their ability to be directly embedded in the web interface of our ERP.  Other open-source solutions that could have been used instead of Grafana include SuperSet~\cite{superset} and Metabase~\cite{metabase}.

\subsection{AI development and analysis}
The application of AI in ITOps requires extensive exploration and experimentation to identify effective methods for addressing related challenges. As such, it is important to offer an efficient collaborative framework and toolset for data-scientists. This starts with providing a Version Control System to manage and version all source code related to experimental studies and development. Our team exploits GitLab~\cite{gitlab}, an open-source git platform that can be used for version control, CI/CD, automatic unit testing, and more. We use Python as a programming language as it is the language of choice for data science, but also, it is completely apt and well-equipped to run in production. We host a Jupyter Hub~\cite{jupyterHub} to facilitate collaborative work around Python notebooks. Each data scientist can also benefit from using Visual Studio as an IDE, which offers integrated features such as auto-completion, git feature simplification, code formatting, and error detection. To perfectly isolate between different AI projects, a separate Python virtual environment is employed for each project with Poetry~\cite{poetry} as a virtual environment manager.

\subsection{AI in Production}
Once effective algorithms have been developed and tested, they can be transitioned into production. Our infrastructure supports the deployment of both HTTP RESTful query-response services and scheduled data transformation tasks. FastAPI~\cite{fastapi} has emerged as a popular choice for managing Python services in production environments. By using Nginx~\cite{nginx} as a proxy, we can redirect HTTP queries to the appropriate service, thereby facilitating the deployment of multiple services. Docker containers can be used to isolate services and streamline their deployment. Conversely, certain Python routines, such as data transformations (ETL), are more appropriately defined as tasks. These tasks are orchestrated using a built-in job scheduler in our ERP system, a tool that is also widely employed in several other modules of Copilote. Open-source alternatives for orchestrating Python jobs include Airflow~\cite{airflow} and Prefect~\cite{prefect}. Python tasks can be scheduled based on specific frequency or triggered by certain events. Orchestration tools also allow for the management of dependent job executions, where the execution of one job is contingent on the output of others.

\subsection{Alert engine}

Alerting is a core feature of predictive maintenance. It aims to generate timely alerts when anomalies or other relevant issues are detected (resp. predicted) in the current (resp. future) state of the system. These alerts can be created through both simple threshold-based rules and sophisticated ML techniques. To enable a full integration with our AIOps module and internal incident management tool, we have developed our own alert engine. This engine can access ERP data via Copilote Infocenter and can interact with external databases like ClickHouse. Additionally, it can trigger alerts based on computations produced by an AI service. This engine can be used to create, update, and fine-tune alerts. The definition of an alert needs to be expressed with a simple rule such as:  \texttt{if metric > threshold} $\implies$ \texttt{generate alert with a given severity level}. This straightforward format is equally effective for complex alerts based on ML methods. For example, the \texttt{metric} used to trigger the alert could be a system anomaly score, computed beforehand with an LSTM auto-encoder~\cite{nedelkoski2019anomaly}. One of our usage of ML methods is time series forecasting to predict the time remained before saturation of resources such as storage, memory, swap, and Oracle table space. We have employed a hybrid approach by combining simple Linear regression and a Deep Neural network approach called NHITS~\cite{DBLP:conf/aaai/ChalluOORCD23} which has shown its efficiency in long horizon time series forecasting. These methods are used to predict the number of remaining days before saturation, then, a simple alert rule is configured when this metric is lower than a specific threshold. Moreover, alert rules can be configured to initiate automatic actions, such as creating and assigning an incident ticket or applying a corrective action. Several open-source solutions, like Grafana~\cite{grafana} and Kibana~\cite{kibana}, can also be used as alert engines. While Grafana has the capability to define alerts on various data sources, Kibana is limited only to Elasticsearch. It's worth noting that using such products involves the additional cost of interfacing alerts with our own system.

\begin{figure}[t]
	\centering
	\includegraphics[width=0.95\linewidth]{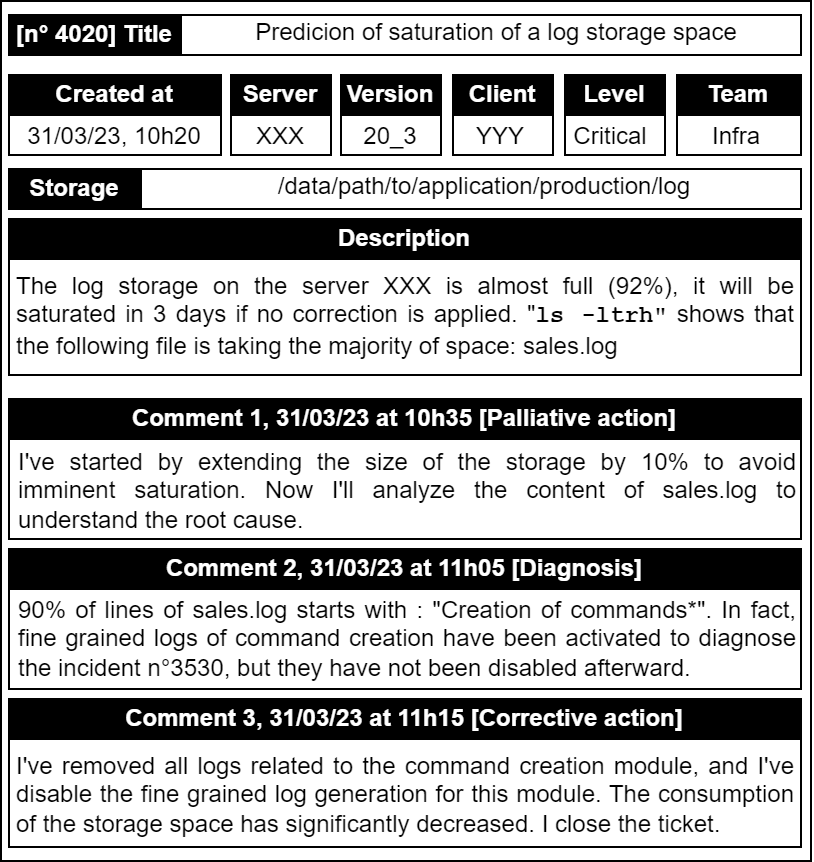}
    \captionsetup{font=footnotesize}
	\caption{Example of an incident ticket.}%
	\label{fig:incidentTicket}
\end{figure}

\subsection{Incident Management and resolution}
After a problem has been detected, an incident ticket is created to address this problem. This ticket can be created automatically with the alert engine, or manually by a Copilote user. An incident includes a title and description, as well as contextualized attributes to identify the concerned server, application, client, the time of incident, etc., as shown in Fig~\ref{fig:incidentTicket}. Managing incidents includes several operations. It first starts by assigning the incident to the right team or person, as maintenance staff are organized on several teams that are specialized on different topics (database, application, infrastructure, etc.). This is called incident triage and it can be completely automated with efficient ML methods~\cite{DBLP:conf/icse/0003HLXZHGXDZ19,DBLP:conf/kbse/0003HL0HGXDZ19,DBLP:conf/dsaa/RemilBPRK21}. Additionally, incidents of each team need to be ranked with respect to their criticality. Then, for each incident, the assigned person starts by diagnosing the problem in order to understand its origin. It is noteworthy that in some urgent cases, the person starts by applying a palliative action to unlock the problem. To make the diagnosis process efficient, maintenance teams need to be equipped with effective tools that enable them to contextualize incidents and perform root cause analysis. Once the root cause is identified, corrective actions can be applied. The incident ticket also contains a list of timestamped comments that explain the operations that have been done to diagnose and solve the problem, as illustrated in Fig~\ref{fig:incidentTicket}. In our Copilote AIOps module, we have implemented an incident management node that covers all the aforementioned operations and automate as many steps as possible. Alternatively, one could exploit iTop~\cite{itop} which is an open-source incident management system.

\section{Conclusion\label{sec:discussion}} 

Setting up an on-premise AIOps architecture is a strategic choice for medium-size software engineering companies when it comes to keep costs under control but also data. In this article, we presented our AIOps architecture and most importantly what led us there according to a set of adaptable criteria to guide decision-making, including data sovereignty, cost controls, ease of use, adaptability and stability among others. We believe that our methodology can be adapted to many companies, but also simply brings elements of discussion in general. Shifting it towards the cloud is possible and is, again, a strategic decision\footnote{See for example \url{https://world.hey.com/dhh/why-we-re-leaving-the-cloud-654b47e0}}.

We now run our approach in production. We drastically reduced the number of fatal incidents of all the components we monitor. For example, we almost eradicated Copilote service unavailability's due to disks/databases saturation, JVM memory leaks, UNIX swap leaks, and many more. Our infrastructure is stable: the only components we changed since the first production launches (we proceeded iteratively and added new features each time) are the NoSQL databases.  The cost of handling our infrastructure notably decreased when we moved from using two solutions (InfluxDB and Elasticsearch) for our Data LakeHouse to only one (ClickHouse) and it came with a better adoption as \textit{anyone knows SQL}. It is also easy to set up new data collections and alerts based on rules: anyone knowing either SQL or the Copilote BI module can set up an alert. For predictive and anomaly detection tasks, we still need dedicated studies and programs, generally in Python using well-known libraries such that scikit-learn~\cite{scikit-learn} or specialized algorithms \cite{DBLP:conf/aaai/ChalluOORCD23}. However, we are actively researching more efficient tools to facilitate the integration of ML/DM, including model and feature stores, as well as specialized tools for monitoring models in production (see e.g. MindsDB). Our second main line of research is to explore the application of Large Language Models~\cite{DBLP:conf/nips/VaswaniSPUJGKP17,DBLP:conf/acl/LewisLGGMLSZ20,DBLP:conf/naacl/DevlinCLT19,DBLP:journals/mima/FloridiC20,DBLP:journals/corr/abs-2303-12712} in various aspects of maintenance. We can analyze historical incident tickets to automatically diagnose new incidents, provide recommended corrections, and generate summaries. Currently, ChatGPT~\cite{chatgpt}, a model owned by OpenAI, is remarkably more accurate than open-source alternatives. However, we are witnessing rapid progress in open-source LLMs~\cite{openLlmLeaderboard} that can be trained and deployed on-premise, avoiding the need to share private data to third parties.

\bibliographystyle{ACM-Reference-Format}
\bibliography{biblio}


\begin{thebibliography}{68}


\ifx \showCODEN    \undefined \def \showCODEN     #1{\unskip}     \fi
\ifx \showDOI      \undefined \def \showDOI       #1{#1}\fi
\ifx \showISBNx    \undefined \def \showISBNx     #1{\unskip}     \fi
\ifx \showISBNxiii \undefined \def \showISBNxiii  #1{\unskip}     \fi
\ifx \showISSN     \undefined \def \showISSN      #1{\unskip}     \fi
\ifx \showLCCN     \undefined \def \showLCCN      #1{\unskip}     \fi
\ifx \shownote     \undefined \def \shownote      #1{#1}          \fi
\ifx \showarticletitle \undefined \def \showarticletitle #1{#1}   \fi
\ifx \showURL      \undefined \def \showURL       {\relax}        \fi
\providecommand\bibfield[2]{#2}
\providecommand\bibinfo[2]{#2}
\providecommand\natexlab[1]{#1}
\providecommand\showeprint[2][]{arXiv:#2}

\bibitem[ans(2023)]%
        {ansible}
 \bibinfo{year}{2023}\natexlab{}.
\newblock \bibinfo{title}{Ansible}.
\newblock
\newblock
\urldef\tempurl%
\url{https://www.ansible.com/}
\showURL{%
\tempurl}


\bibitem[air(2023)]%
        {airflow}
 \bibinfo{year}{2023}\natexlab{}.
\newblock \bibinfo{title}{Apache Airflow}.
\newblock
\newblock
\urldef\tempurl%
\url{https://airflow.apache.org/}
\showURL{%
\tempurl}


\bibitem[kaf(2023)]%
        {kafka}
 \bibinfo{year}{2023}\natexlab{}.
\newblock \bibinfo{title}{Apache Kafka}.
\newblock
\newblock
\urldef\tempurl%
\url{https://kafka.apache.org/}
\showURL{%
\tempurl}


\bibitem[sup(2023)]%
        {superset}
 \bibinfo{year}{2023}\natexlab{}.
\newblock \bibinfo{title}{Apache Superset}.
\newblock
\newblock
\urldef\tempurl%
\url{https://superset.apache.org/}
\showURL{%
\tempurl}


\bibitem[bea(2023)]%
        {beats}
 \bibinfo{year}{2023}\natexlab{}.
\newblock \bibinfo{title}{Beats from the ELK stack.}
\newblock
\newblock
\urldef\tempurl%
\url{https://www.elastic.co/fr/beats/}
\showURL{%
\tempurl}


\bibitem[cha(2023)]%
        {chatgpt}
 \bibinfo{year}{2023}\natexlab{}.
\newblock \bibinfo{title}{ChatGPT, the OpenAI chat bot.}
\newblock
\newblock
\urldef\tempurl%
\url{https://chat.openai.com/}
\showURL{%
\tempurl}


\bibitem[cli(2023)]%
        {clickbench}
 \bibinfo{year}{2023}\natexlab{}.
\newblock \bibinfo{title}{ClickBench: a Benchmark For Analytical Databases}.
\newblock
\newblock
\urldef\tempurl%
\url{https://github.com/ClickHouse/ClickBench}
\showURL{%
\tempurl}


\bibitem[con(2023)]%
        {contentsquareClickHouse}
 \bibinfo{year}{2023}\natexlab{}.
\newblock \bibinfo{title}{Contentsquare Migration from Elasticsearch to
  ClickHouse}.
\newblock
\newblock
\urldef\tempurl%
\url{https://clickhouse.com/blog/contentsquare-migration-from-elasticsearch-to-clickhouse}
\showURL{%
\tempurl}


\bibitem[dat(2023)]%
        {datadog}
 \bibinfo{year}{2023}\natexlab{}.
\newblock \bibinfo{title}{Datadog}.
\newblock
\newblock
\urldef\tempurl%
\url{https://www.datadoghq.com/}
\showURL{%
\tempurl}


\bibitem[dis(2023)]%
        {disneyClickHouse}
 \bibinfo{year}{2023}\natexlab{}.
\newblock \bibinfo{title}{Disney+ClickHouse, Disney's Flexible ELT Pipelines in
  ClickHouse}.
\newblock
\newblock
\urldef\tempurl%
\url{https://clickhouse.com/blog/nyc-meetup-report-high-speed-content-distribution-analytics-for-streaming-platforms}
\showURL{%
\tempurl}


\bibitem[dyn(2023)]%
        {dynatrace}
 \bibinfo{year}{2023}\natexlab{}.
\newblock \bibinfo{title}{Dynatrace}.
\newblock
\newblock
\urldef\tempurl%
\url{https://www.dynatrace.com/}
\showURL{%
\tempurl}


\bibitem[ela(2023)]%
        {elasticsearch}
 \bibinfo{year}{2023}\natexlab{}.
\newblock \bibinfo{title}{Elasticsearch}.
\newblock
\newblock
\urldef\tempurl%
\url{https://www.elastic.co/}
\showURL{%
\tempurl}


\bibitem[fas(2023)]%
        {fastapi}
 \bibinfo{year}{2023}\natexlab{}.
\newblock \bibinfo{title}{FastAPI}.
\newblock
\newblock
\urldef\tempurl%
\url{https://fastapi.tiangolo.com/}
\showURL{%
\tempurl}


\bibitem[flu(2023)]%
        {fluentd}
 \bibinfo{year}{2023}\natexlab{}.
\newblock \bibinfo{title}{Fluentd}.
\newblock
\newblock
\urldef\tempurl%
\url{https://www.fluentd.org/}
\showURL{%
\tempurl}


\bibitem[gdp(2023)]%
        {gdpr}
 \bibinfo{year}{2023}\natexlab{}.
\newblock \bibinfo{title}{GDPR}.
\newblock
\newblock
\urldef\tempurl%
\url{https://gdpr.eu/}
\showURL{%
\tempurl}


\bibitem[git(2023)]%
        {gitlab}
 \bibinfo{year}{2023}\natexlab{}.
\newblock \bibinfo{title}{GitLab}.
\newblock
\newblock
\urldef\tempurl%
\url{https://gitlab.com/gitlab-org/gitlab}
\showURL{%
\tempurl}


\bibitem[gra(2023)]%
        {grafana}
 \bibinfo{year}{2023}\natexlab{}.
\newblock \bibinfo{title}{Grafana}.
\newblock
\newblock
\urldef\tempurl%
\url{https://grafana.com/}
\showURL{%
\tempurl}


\bibitem[clo(2023)]%
        {cloudfareClickHouse}
 \bibinfo{year}{2023}\natexlab{}.
\newblock \bibinfo{title}{HTTP Analytics for 6M requests per second using
  ClickHouse}.
\newblock
\newblock
\urldef\tempurl%
\url{https://blog.cloudflare.com/http-analytics-for-6m-requests-per-second-using-clickhouse/}
\showURL{%
\tempurl}


\bibitem[ope(2023)]%
        {openLlmLeaderboard}
 \bibinfo{year}{2023}\natexlab{}.
\newblock \bibinfo{title}{HuggingFace Open LLM Leaderboard}.
\newblock
\newblock
\urldef\tempurl%
\url{https://huggingface.co/spaces/HuggingFaceH4/open_llm_leaderboard}
\showURL{%
\tempurl}


\bibitem[inf(2023)]%
        {infologiccopilote}
 \bibinfo{year}{2023}\natexlab{}.
\newblock \bibinfo{title}{Infologic-Copilote}.
\newblock
\newblock
\urldef\tempurl%
\url{https://www.infologic-copilote.fr/}
\showURL{%
\tempurl}


\bibitem[ins(2023)]%
        {instana}
 \bibinfo{year}{2023}\natexlab{}.
\newblock \bibinfo{title}{Instana}.
\newblock
\newblock
\urldef\tempurl%
\url{https://www.instana.com/}
\showURL{%
\tempurl}


\bibitem[ito(2023)]%
        {itop}
 \bibinfo{year}{2023}\natexlab{}.
\newblock \bibinfo{title}{iTop}.
\newblock
\newblock
\urldef\tempurl%
\url{https://www.combodo.com/itop-193}
\showURL{%
\tempurl}


\bibitem[jup(2023)]%
        {jupyterHub}
 \bibinfo{year}{2023}\natexlab{}.
\newblock \bibinfo{title}{JupyterHub}.
\newblock
\newblock
\urldef\tempurl%
\url{https://jupyter.org/hub}
\showURL{%
\tempurl}


\bibitem[kib(2023)]%
        {kibana}
 \bibinfo{year}{2023}\natexlab{}.
\newblock \bibinfo{title}{Kibana from the ELK stack}.
\newblock
\newblock
\urldef\tempurl%
\url{https://www.elastic.co/fr/kibana/}
\showURL{%
\tempurl}


\bibitem[log(2023)]%
        {logstach}
 \bibinfo{year}{2023}\natexlab{}.
\newblock \bibinfo{title}{Logstash}.
\newblock
\newblock
\urldef\tempurl%
\url{https://www.elastic.co/fr/logstash/}
\showURL{%
\tempurl}


\bibitem[met(2023)]%
        {metabase}
 \bibinfo{year}{2023}\natexlab{}.
\newblock \bibinfo{title}{Metabase}.
\newblock
\newblock
\urldef\tempurl%
\url{https://www.metabase.com/}
\showURL{%
\tempurl}


\bibitem[mon(2023)]%
        {mongodb}
 \bibinfo{year}{2023}\natexlab{}.
\newblock \bibinfo{title}{MongoDB}.
\newblock
\newblock
\urldef\tempurl%
\url{https://www.mongodb.com/}
\showURL{%
\tempurl}


\bibitem[new(2023)]%
        {newrelic}
 \bibinfo{year}{2023}\natexlab{}.
\newblock \bibinfo{title}{New Relic}.
\newblock
\newblock
\urldef\tempurl%
\url{https://newrelic.com/fr}
\showURL{%
\tempurl}


\bibitem[ngi(2023)]%
        {nginx}
 \bibinfo{year}{2023}\natexlab{}.
\newblock \bibinfo{title}{Nginx}.
\newblock
\newblock
\urldef\tempurl%
\url{https://www.nginx.com/}
\showURL{%
\tempurl}


\bibitem[eba(2023)]%
        {ebayClickHouse}
 \bibinfo{year}{2023}\natexlab{}.
\newblock \bibinfo{title}{Our Online Analytical Processing Journey with
  ClickHouse on Kubernetes}.
\newblock
\newblock
\urldef\tempurl%
\url{https://tech.ebayinc.com/engineering/ou-online-analytical-processing/}
\showURL{%
\tempurl}


\bibitem[pag(2023)]%
        {pagerduty}
 \bibinfo{year}{2023}\natexlab{}.
\newblock \bibinfo{title}{{PagerDuty}}.
\newblock
\newblock
\urldef\tempurl%
\url{https://www.pagerduty.com/}
\showURL{%
\tempurl}


\bibitem[poe(2023)]%
        {poetry}
 \bibinfo{year}{2023}\natexlab{}.
\newblock \bibinfo{title}{Poetry}.
\newblock
\newblock
\urldef\tempurl%
\url{https://python-poetry.org/}
\showURL{%
\tempurl}


\bibitem[pre(2023)]%
        {prefect}
 \bibinfo{year}{2023}\natexlab{}.
\newblock \bibinfo{title}{Prefect}.
\newblock
\newblock
\urldef\tempurl%
\url{https://www.prefect.io/}
\showURL{%
\tempurl}


\bibitem[rab(2023)]%
        {rabbitmq}
 \bibinfo{year}{2023}\natexlab{}.
\newblock \bibinfo{title}{RabbitMQ}.
\newblock
\newblock
\urldef\tempurl%
\url{https://www.rabbitmq.com/}
\showURL{%
\tempurl}


\bibitem[roc(2023)]%
        {rocketmq}
 \bibinfo{year}{2023}\natexlab{}.
\newblock \bibinfo{title}{RocketMQ}.
\newblock
\newblock
\urldef\tempurl%
\url{https://rocketmq.apache.org/}
\showURL{%
\tempurl}


\bibitem[spl(2023)]%
        {splunk}
 \bibinfo{year}{2023}\natexlab{}.
\newblock \bibinfo{title}{Splunk}.
\newblock
\newblock
\urldef\tempurl%
\url{https://www.splunk.com/}
\showURL{%
\tempurl}


\bibitem[tel(2023)]%
        {telegraf}
 \bibinfo{year}{2023}\natexlab{}.
\newblock \bibinfo{title}{Telegraf from InfluxData.}
\newblock
\newblock
\urldef\tempurl%
\url{https://www.influxdata.com/time-series-platform/telegraf/}
\showURL{%
\tempurl}


\bibitem[Bogatinovski et~al\mbox{.}(2021)]%
        {bogatinovski2021artificial}
\bibfield{author}{\bibinfo{person}{Jasmin Bogatinovski}, \bibinfo{person}{Sasho
  Nedelkoski}, \bibinfo{person}{Alexander Acker}, \bibinfo{person}{Florian
  Schmidt}, \bibinfo{person}{Thorsten Wittkopp}, \bibinfo{person}{Soeren
  Becker}, \bibinfo{person}{Jorge Cardoso}, {and} \bibinfo{person}{Odej Kao}.}
  \bibinfo{year}{2021}\natexlab{}.
\newblock \showarticletitle{Artificial Intelligence for IT Operations (AIOPS)
  Workshop White Paper}.
\newblock \bibinfo{journal}{\emph{arXiv preprint arXiv:2101.06054}}
  (\bibinfo{year}{2021}).
\newblock


\bibitem[Bubeck et~al\mbox{.}(2023)]%
        {DBLP:journals/corr/abs-2303-12712}
\bibfield{author}{\bibinfo{person}{S{\'{e}}bastien Bubeck},
  \bibinfo{person}{Varun Chandrasekaran}, \bibinfo{person}{Ronen Eldan},
  \bibinfo{person}{Johannes Gehrke}, \bibinfo{person}{Eric Horvitz},
  \bibinfo{person}{Ece Kamar}, \bibinfo{person}{Peter Lee},
  \bibinfo{person}{Yin~Tat Lee}, \bibinfo{person}{Yuanzhi Li},
  \bibinfo{person}{Scott~M. Lundberg}, \bibinfo{person}{Harsha Nori},
  \bibinfo{person}{Hamid Palangi}, \bibinfo{person}{Marco~T{\'{u}}lio Ribeiro},
  {and} \bibinfo{person}{Yi Zhang}.} \bibinfo{year}{2023}\natexlab{}.
\newblock \showarticletitle{Sparks of Artificial General Intelligence: Early
  experiments with {GPT-4}}.
\newblock \bibinfo{journal}{\emph{CoRR}}  \bibinfo{volume}{abs/2303.12712}
  (\bibinfo{year}{2023}).
\newblock


\bibitem[Challu et~al\mbox{.}(2023)]%
        {DBLP:conf/aaai/ChalluOORCD23}
\bibfield{author}{\bibinfo{person}{Cristian Challu}, \bibinfo{person}{Kin~G.
  Olivares}, \bibinfo{person}{Boris~N. Oreshkin},
  \bibinfo{person}{Federico~Garza Ram{\'{\i}}rez},
  \bibinfo{person}{Max~Mergenthaler Canseco}, {and} \bibinfo{person}{Artur
  Dubrawski}.} \bibinfo{year}{2023}\natexlab{}.
\newblock \showarticletitle{{NHITS:} Neural Hierarchical Interpolation for Time
  Series Forecasting}. In \bibinfo{booktitle}{\emph{Thirty-Seventh {AAAI}
  Conference on Artificial Intelligence, {AAAI} 2023, Thirty-Fifth Conference
  on Innovative Applications of Artificial Intelligence, {IAAI} 2023,
  Thirteenth Symposium on Educational Advances in Artificial Intelligence,
  {EAAI} 2023, Washington, DC, USA, February 7-14, 2023}},
  \bibfield{editor}{\bibinfo{person}{Brian Williams}, \bibinfo{person}{Yiling
  Chen}, {and} \bibinfo{person}{Jennifer Neville}} (Eds.).
  \bibinfo{publisher}{{AAAI} Press}, \bibinfo{pages}{6989--6997}.
\newblock


\bibitem[Chen et~al\mbox{.}(2019a)]%
        {DBLP:conf/icse/0003HLXZHGXDZ19}
\bibfield{author}{\bibinfo{person}{Junjie Chen}, \bibinfo{person}{Xiaoting He},
  \bibinfo{person}{Qingwei Lin}, \bibinfo{person}{Yong Xu},
  \bibinfo{person}{Hongyu Zhang}, \bibinfo{person}{Dan Hao},
  \bibinfo{person}{Feng Gao}, \bibinfo{person}{Zhangwei Xu},
  \bibinfo{person}{Yingnong Dang}, {and} \bibinfo{person}{Dongmei Zhang}.}
  \bibinfo{year}{2019}\natexlab{a}.
\newblock \showarticletitle{An empirical investigation of incident triage for
  online service systems}. In \bibinfo{booktitle}{\emph{Proceedings of the 41st
  International Conference on Software Engineering: Software Engineering in
  Practice, {ICSE} {(SEIP)} 2019, Montreal, QC, Canada, May 25-31, 2019}},
  \bibfield{editor}{\bibinfo{person}{Helen Sharp} {and} \bibinfo{person}{Mike
  Whalen}} (Eds.). \bibinfo{publisher}{{IEEE} / {ACM}},
  \bibinfo{pages}{111--120}.
\newblock


\bibitem[Chen et~al\mbox{.}(2019b)]%
        {DBLP:conf/kbse/0003HL0HGXDZ19}
\bibfield{author}{\bibinfo{person}{Junjie Chen}, \bibinfo{person}{Xiaoting He},
  \bibinfo{person}{Qingwei Lin}, \bibinfo{person}{Hongyu Zhang},
  \bibinfo{person}{Dan Hao}, \bibinfo{person}{Feng Gao},
  \bibinfo{person}{Zhangwei Xu}, \bibinfo{person}{Yingnong Dang}, {and}
  \bibinfo{person}{Dongmei Zhang}.} \bibinfo{year}{2019}\natexlab{b}.
\newblock \showarticletitle{Continuous Incident Triage for Large-Scale Online
  Service Systems}. In \bibinfo{booktitle}{\emph{34th {IEEE/ACM} International
  Conference on Automated Software Engineering, {ASE} 2019, San Diego, CA, USA,
  November 11-15, 2019}}. \bibinfo{publisher}{{IEEE}},
  \bibinfo{pages}{364--375}.
\newblock


\bibitem[Dang et~al\mbox{.}(2019)]%
        {dang2019aiops}
\bibfield{author}{\bibinfo{person}{Yingnong Dang}, \bibinfo{person}{Qingwei
  Lin}, {and} \bibinfo{person}{Peng Huang}.} \bibinfo{year}{2019}\natexlab{}.
\newblock \showarticletitle{AIOps: real-world challenges and research
  innovations}. In \bibinfo{booktitle}{\emph{2019 IEEE/ACM 41st International
  Conference on Software Engineering: Companion Proceedings (ICSE-Companion)}}.
  IEEE, \bibinfo{pages}{4--5}.
\newblock


\bibitem[Devlin et~al\mbox{.}(2019)]%
        {DBLP:conf/naacl/DevlinCLT19}
\bibfield{author}{\bibinfo{person}{Jacob Devlin}, \bibinfo{person}{Ming{-}Wei
  Chang}, \bibinfo{person}{Kenton Lee}, {and} \bibinfo{person}{Kristina
  Toutanova}.} \bibinfo{year}{2019}\natexlab{}.
\newblock \showarticletitle{{BERT:} Pre-training of Deep Bidirectional
  Transformers for Language Understanding}. In
  \bibinfo{booktitle}{\emph{Proceedings of the 2019 Conference of the North
  American Chapter of the Association for Computational Linguistics: Human
  Language Technologies, {NAACL-HLT} 2019, Minneapolis, MN, USA, June 2-7,
  2019, Volume 1 (Long and Short Papers)}},
  \bibfield{editor}{\bibinfo{person}{Jill Burstein}, \bibinfo{person}{Christy
  Doran}, {and} \bibinfo{person}{Thamar Solorio}} (Eds.).
  \bibinfo{publisher}{Association for Computational Linguistics},
  \bibinfo{pages}{4171--4186}.
\newblock


\bibitem[Farshchi et~al\mbox{.}(2018)]%
        {farshchi2018metric}
\bibfield{author}{\bibinfo{person}{Mostafa Farshchi}, \bibinfo{person}{Jean-Guy
  Schneider}, \bibinfo{person}{Ingo Weber}, {and} \bibinfo{person}{John
  Grundy}.} \bibinfo{year}{2018}\natexlab{}.
\newblock \showarticletitle{Metric selection and anomaly detection for cloud
  operations using log and metric correlation analysis}.
\newblock \bibinfo{journal}{\emph{Journal of Systems and Software}}
  \bibinfo{volume}{137} (\bibinfo{year}{2018}), \bibinfo{pages}{531--549}.
\newblock


\bibitem[Floridi and Chiriatti(2020)]%
        {DBLP:journals/mima/FloridiC20}
\bibfield{author}{\bibinfo{person}{Luciano Floridi} {and}
  \bibinfo{person}{Massimo Chiriatti}.} \bibinfo{year}{2020}\natexlab{}.
\newblock \showarticletitle{{GPT-3:} Its Nature, Scope, Limits, and
  Consequences}.
\newblock \bibinfo{journal}{\emph{Minds Mach.}} \bibinfo{volume}{30},
  \bibinfo{number}{4} (\bibinfo{year}{2020}), \bibinfo{pages}{681--694}.
\newblock


\bibitem[Fu et~al\mbox{.}(2021)]%
        {DBLP:journals/access/FuZY21}
\bibfield{author}{\bibinfo{person}{Guo Fu}, \bibinfo{person}{Yanfeng Zhang},
  {and} \bibinfo{person}{Ge Yu}.} \bibinfo{year}{2021}\natexlab{}.
\newblock \showarticletitle{A Fair Comparison of Message Queuing Systems}.
\newblock \bibinfo{journal}{\emph{{IEEE} Access}}  \bibinfo{volume}{9}
  (\bibinfo{year}{2021}), \bibinfo{pages}{421--432}.
\newblock


\bibitem[Harby and Zulkernine(2022)]%
        {harby2022data}
\bibfield{author}{\bibinfo{person}{Ahmed~A Harby} {and}
  \bibinfo{person}{Farhana Zulkernine}.} \bibinfo{year}{2022}\natexlab{}.
\newblock \showarticletitle{From Data Warehouse to Lakehouse: A Comparative
  Review}. In \bibinfo{booktitle}{\emph{2022 IEEE International Conference on
  Big Data (Big Data)}}. IEEE, \bibinfo{pages}{389--395}.
\newblock


\bibitem[Kreuzberger et~al\mbox{.}(2023)]%
        {kreuzberger2023machine}
\bibfield{author}{\bibinfo{person}{Dominik Kreuzberger},
  \bibinfo{person}{Niklas K{\"u}hl}, {and} \bibinfo{person}{Sebastian
  Hirschl}.} \bibinfo{year}{2023}\natexlab{}.
\newblock \showarticletitle{Machine learning operations (mlops): Overview,
  definition, and architecture}.
\newblock \bibinfo{journal}{\emph{IEEE Access}} (\bibinfo{year}{2023}).
\newblock


\bibitem[Lewis et~al\mbox{.}(2020)]%
        {DBLP:conf/acl/LewisLGGMLSZ20}
\bibfield{author}{\bibinfo{person}{Mike Lewis}, \bibinfo{person}{Yinhan Liu},
  \bibinfo{person}{Naman Goyal}, \bibinfo{person}{Marjan Ghazvininejad},
  \bibinfo{person}{Abdelrahman Mohamed}, \bibinfo{person}{Omer Levy},
  \bibinfo{person}{Veselin Stoyanov}, {and} \bibinfo{person}{Luke
  Zettlemoyer}.} \bibinfo{year}{2020}\natexlab{}.
\newblock \showarticletitle{{BART:} Denoising Sequence-to-Sequence Pre-training
  for Natural Language Generation, Translation, and Comprehension}. In
  \bibinfo{booktitle}{\emph{Proceedings of the 58th Annual Meeting of the
  Association for Computational Linguistics, {ACL} 2020, Online, July 5-10,
  2020}}, \bibfield{editor}{\bibinfo{person}{Dan Jurafsky},
  \bibinfo{person}{Joyce Chai}, \bibinfo{person}{Natalie Schluter}, {and}
  \bibinfo{person}{Joel~R. Tetreault}} (Eds.). \bibinfo{publisher}{Association
  for Computational Linguistics}, \bibinfo{pages}{7871--7880}.
\newblock


\bibitem[Lu et~al\mbox{.}(2018)]%
        {lu2018learning}
\bibfield{author}{\bibinfo{person}{Jie Lu}, \bibinfo{person}{Anjin Liu},
  \bibinfo{person}{Fan Dong}, \bibinfo{person}{Feng Gu}, \bibinfo{person}{Joao
  Gama}, {and} \bibinfo{person}{Guangquan Zhang}.}
  \bibinfo{year}{2018}\natexlab{}.
\newblock \showarticletitle{Learning under concept drift: A review}.
\newblock \bibinfo{journal}{\emph{IEEE transactions on knowledge and data
  engineering}} \bibinfo{volume}{31}, \bibinfo{number}{12}
  (\bibinfo{year}{2018}), \bibinfo{pages}{2346--2363}.
\newblock


\bibitem[Lyu et~al\mbox{.}(2021)]%
        {lyu2021towards}
\bibfield{author}{\bibinfo{person}{Yingzhe Lyu}, \bibinfo{person}{Gopi~Krishnan
  Rajbahadur}, \bibinfo{person}{Dayi Lin}, \bibinfo{person}{Boyuan Chen}, {and}
  \bibinfo{person}{Zhen~Ming Jiang}.} \bibinfo{year}{2021}\natexlab{}.
\newblock \showarticletitle{Towards a consistent interpretation of aiops
  models}.
\newblock \bibinfo{journal}{\emph{ACM Transactions on Software Engineering and
  Methodology (TOSEM)}} \bibinfo{volume}{31}, \bibinfo{number}{1}
  (\bibinfo{year}{2021}), \bibinfo{pages}{1--38}.
\newblock


\bibitem[Mostafa et~al\mbox{.}(2022)]%
        {mostafa2022scits}
\bibfield{author}{\bibinfo{person}{Jalal Mostafa}, \bibinfo{person}{Sara
  Wehbi}, \bibinfo{person}{Suren Chilingaryan}, {and} \bibinfo{person}{Andreas
  Kopmann}.} \bibinfo{year}{2022}\natexlab{}.
\newblock \showarticletitle{SciTS: A Benchmark for Time-Series Databases in
  Scientific Experiments and Industrial Internet of Things}. In
  \bibinfo{booktitle}{\emph{Proceedings of the 34th International Conference on
  Scientific and Statistical Database Management}}. \bibinfo{pages}{1--11}.
\newblock


\bibitem[Nedelkoski et~al\mbox{.}(2019)]%
        {nedelkoski2019anomaly}
\bibfield{author}{\bibinfo{person}{Sasho Nedelkoski}, \bibinfo{person}{Jorge
  Cardoso}, {and} \bibinfo{person}{Odej Kao}.} \bibinfo{year}{2019}\natexlab{}.
\newblock \showarticletitle{Anomaly detection from system tracing data using
  multimodal deep learning}. In \bibinfo{booktitle}{\emph{2019 IEEE 12th
  International Conference on Cloud Computing (CLOUD)}}. IEEE,
  \bibinfo{pages}{179--186}.
\newblock


\bibitem[Notaro et~al\mbox{.}(2021)]%
        {notaro2021systematic}
\bibfield{author}{\bibinfo{person}{Paolo Notaro}, \bibinfo{person}{Jorge
  Cardoso}, {and} \bibinfo{person}{Michael Gerndt}.}
  \bibinfo{year}{2021}\natexlab{}.
\newblock \showarticletitle{A systematic mapping study in AIOps}. In
  \bibinfo{booktitle}{\emph{Service-Oriented Computing--ICSOC 2020 Workshops:
  AIOps, CFTIC, STRAPS, AI-PA, AI-IOTS, and Satellite Events, Dubai, United
  Arab Emirates, December 14--17, 2020, Proceedings}}. Springer,
  \bibinfo{pages}{110--123}.
\newblock


\bibitem[Pandey and Mittal(2020)]%
        {pandey2020analogy}
\bibfield{author}{\bibinfo{person}{Himanshu Pandey} {and}
  \bibinfo{person}{Er~Kushagra Mittal}.} \bibinfo{year}{2020}\natexlab{}.
\newblock \showarticletitle{Analogy between Agent Less Monitoring and Agent
  Based Monitoring}.
\newblock \bibinfo{journal}{\emph{Reliability: Theory \& Applications}}
  \bibinfo{volume}{15}, \bibinfo{number}{3} (\bibinfo{year}{2020}),
  \bibinfo{pages}{117--124}.
\newblock


\bibitem[Pedregosa et~al\mbox{.}(2011)]%
        {scikit-learn}
\bibfield{author}{\bibinfo{person}{F. Pedregosa}, \bibinfo{person}{G.
  Varoquaux}, \bibinfo{person}{A. Gramfort}, \bibinfo{person}{V. Michel},
  \bibinfo{person}{B. Thirion}, \bibinfo{person}{O. Grisel},
  \bibinfo{person}{M. Blondel}, \bibinfo{person}{P. Prettenhofer},
  \bibinfo{person}{R. Weiss}, \bibinfo{person}{V. Dubourg}, \bibinfo{person}{J.
  Vanderplas}, \bibinfo{person}{A. Passos}, \bibinfo{person}{D. Cournapeau},
  \bibinfo{person}{M. Brucher}, \bibinfo{person}{M. Perrot}, {and}
  \bibinfo{person}{E. Duchesnay}.} \bibinfo{year}{2011}\natexlab{}.
\newblock \showarticletitle{Scikit-learn: Machine Learning in {P}ython}.
\newblock \bibinfo{journal}{\emph{Journal of Machine Learning Research}}
  \bibinfo{volume}{12} (\bibinfo{year}{2011}), \bibinfo{pages}{2825--2830}.
\newblock


\bibitem[Prasad and Rich(2018)]%
        {prasad2018market}
\bibfield{author}{\bibinfo{person}{Pankaj Prasad} {and}
  \bibinfo{person}{Charley Rich}.} \bibinfo{year}{2018}\natexlab{}.
\newblock \showarticletitle{Market Guide for AIOps Platforms}.
\newblock \bibinfo{journal}{\emph{Retrieved March}}  \bibinfo{volume}{12}
  (\bibinfo{year}{2018}), \bibinfo{pages}{2020}.
\newblock


\bibitem[Premkumar et~al\mbox{.}(1994)]%
        {DBLP:journals/jmis/PremkumarRN94}
\bibfield{author}{\bibinfo{person}{G.~Prem Premkumar},
  \bibinfo{person}{Keshavamurthy Ramamurthy}, {and} \bibinfo{person}{Sree
  Nilakanta}.} \bibinfo{year}{1994}\natexlab{}.
\newblock \showarticletitle{Implementation of Electronic Data Interchange: An
  Innovation Diffusion Perspective}.
\newblock \bibinfo{journal}{\emph{J. Manag. Inf. Syst.}} \bibinfo{volume}{11},
  \bibinfo{number}{2} (\bibinfo{year}{1994}), \bibinfo{pages}{157--186}.
\newblock


\bibitem[Remil et~al\mbox{.}(2021)]%
        {DBLP:conf/dsaa/RemilBPRK21}
\bibfield{author}{\bibinfo{person}{Youcef Remil}, \bibinfo{person}{Anes
  Bendimerad}, \bibinfo{person}{Marc Plantevit}, \bibinfo{person}{C{\'{e}}line
  Robardet}, {and} \bibinfo{person}{Mehdi Kaytoue}.}
  \bibinfo{year}{2021}\natexlab{}.
\newblock \showarticletitle{Interpretable Summaries of Black Box Incident
  Triaging with Subgroup Discovery}. In \bibinfo{booktitle}{\emph{8th {IEEE}
  International Conference on Data Science and Advanced Analytics, {DSAA} 2021,
  Porto, Portugal, October 6-9, 2021}}. \bibinfo{publisher}{{IEEE}},
  \bibinfo{pages}{1--10}.
\newblock


\bibitem[Ren et~al\mbox{.}(2019)]%
        {ren2019time}
\bibfield{author}{\bibinfo{person}{Hansheng Ren}, \bibinfo{person}{Bixiong Xu},
  \bibinfo{person}{Yujing Wang}, \bibinfo{person}{Chao Yi},
  \bibinfo{person}{Congrui Huang}, \bibinfo{person}{Xiaoyu Kou},
  \bibinfo{person}{Tony Xing}, \bibinfo{person}{Mao Yang}, \bibinfo{person}{Jie
  Tong}, {and} \bibinfo{person}{Qi Zhang}.} \bibinfo{year}{2019}\natexlab{}.
\newblock \showarticletitle{Time-series anomaly detection service at
  microsoft}. In \bibinfo{booktitle}{\emph{Proceedings of the 25th ACM SIGKDD
  International Conference on Knowledge Discovery \& Data Mining}}.
  \bibinfo{pages}{3009--3017}.
\newblock


\bibitem[Ren et~al\mbox{.}(2022)]%
        {DBLP:journals/csur/RenXCHLGCW22}
\bibfield{author}{\bibinfo{person}{Pengzhen Ren}, \bibinfo{person}{Yun Xiao},
  \bibinfo{person}{Xiaojun Chang}, \bibinfo{person}{Po{-}Yao Huang},
  \bibinfo{person}{Zhihui Li}, \bibinfo{person}{Brij~B. Gupta},
  \bibinfo{person}{Xiaojiang Chen}, {and} \bibinfo{person}{Xin Wang}.}
  \bibinfo{year}{2022}\natexlab{}.
\newblock \showarticletitle{A Survey of Deep Active Learning}.
\newblock \bibinfo{journal}{\emph{{ACM} Comput. Surv.}} \bibinfo{volume}{54},
  \bibinfo{number}{9} (\bibinfo{year}{2022}), \bibinfo{pages}{180:1--180:40}.
\newblock


\bibitem[Rijal et~al\mbox{.}(2022)]%
        {rijal2022aiops}
\bibfield{author}{\bibinfo{person}{Laxmi Rijal}, \bibinfo{person}{Ricardo
  Colomo-Palacios}, {and} \bibinfo{person}{Mary S{\'a}nchez-Gord{\'o}n}.}
  \bibinfo{year}{2022}\natexlab{}.
\newblock \showarticletitle{Aiops: A multivocal literature review}.
\newblock \bibinfo{journal}{\emph{Artificial Intelligence for Cloud and Edge
  Computing}} (\bibinfo{year}{2022}), \bibinfo{pages}{31--50}.
\newblock


\bibitem[Settles(2009)]%
        {settles2009active}
\bibfield{author}{\bibinfo{person}{Burr Settles}.}
  \bibinfo{year}{2009}\natexlab{}.
\newblock \showarticletitle{Active learning literature survey}.
\newblock  (\bibinfo{year}{2009}).
\newblock


\bibitem[Shen et~al\mbox{.}(2020)]%
        {shen2020evolving}
\bibfield{author}{\bibinfo{person}{Shijun Shen}, \bibinfo{person}{Jiuling
  Zhang}, \bibinfo{person}{Daochao Huang}, {and} \bibinfo{person}{Jun Xiao}.}
  \bibinfo{year}{2020}\natexlab{}.
\newblock \showarticletitle{Evolving from Traditional Systems to AIOps: Design,
  Implementation and Measurements}. In \bibinfo{booktitle}{\emph{2020 IEEE
  International Conference on Advances in Electrical Engineering and Computer
  Applications (AEECA)}}. IEEE, \bibinfo{pages}{276--280}.
\newblock


\bibitem[Vaswani et~al\mbox{.}(2017)]%
        {DBLP:conf/nips/VaswaniSPUJGKP17}
\bibfield{author}{\bibinfo{person}{Ashish Vaswani}, \bibinfo{person}{Noam
  Shazeer}, \bibinfo{person}{Niki Parmar}, \bibinfo{person}{Jakob Uszkoreit},
  \bibinfo{person}{Llion Jones}, \bibinfo{person}{Aidan~N. Gomez},
  \bibinfo{person}{Lukasz Kaiser}, {and} \bibinfo{person}{Illia Polosukhin}.}
  \bibinfo{year}{2017}\natexlab{}.
\newblock \showarticletitle{Attention is All you Need}. In
  \bibinfo{booktitle}{\emph{Advances in Neural Information Processing Systems
  30: Annual Conference on Neural Information Processing Systems 2017, December
  4-9, 2017, Long Beach, CA, {USA}}},
  \bibfield{editor}{\bibinfo{person}{Isabelle Guyon}, \bibinfo{person}{Ulrike
  von Luxburg}, \bibinfo{person}{Samy Bengio}, \bibinfo{person}{Hanna~M.
  Wallach}, \bibinfo{person}{Rob Fergus}, \bibinfo{person}{S.~V.~N.
  Vishwanathan}, {and} \bibinfo{person}{Roman Garnett}} (Eds.).
  \bibinfo{pages}{5998--6008}.
\newblock


\bibitem[Yeruva(2023)]%
        {yeruva2023monitoring}
\bibfield{author}{\bibinfo{person}{Ajay~Reddy Yeruva}.}
  \bibinfo{year}{2023}\natexlab{}.
\newblock \showarticletitle{Monitoring Data Center Site Infrastructure Using
  AIOPS Architecture}.
\newblock \bibinfo{journal}{\emph{Eduvest-Journal of Universal Studies}}
  \bibinfo{volume}{3}, \bibinfo{number}{1} (\bibinfo{year}{2023}),
  \bibinfo{pages}{265--277}.
\newblock


\bibitem[Zaharia et~al\mbox{.}(2021)]%
        {DBLP:conf/cidr/Zaharia0XA21}
\bibfield{author}{\bibinfo{person}{Matei Zaharia}, \bibinfo{person}{Ali
  Ghodsi}, \bibinfo{person}{Reynold Xin}, {and} \bibinfo{person}{Michael
  Armbrust}.} \bibinfo{year}{2021}\natexlab{}.
\newblock \showarticletitle{Lakehouse: {A} New Generation of Open Platforms
  that Unify Data Warehousing and Advanced Analytics}. In
  \bibinfo{booktitle}{\emph{11th Conference on Innovative Data Systems
  Research, {CIDR} 2021, Virtual Event, January 11-15, 2021, Online
  Proceedings}}. \bibinfo{publisher}{www.cidrdb.org}.
\newblock


\end{thebibliography}

\end{document}